\UseRawInputEncoding
\documentclass[11pt,aps,amssymb,prd,a4paper,nofootinbib]{revtex4-2}
\usepackage{fullpage}
\usepackage{amsfonts}
\usepackage{amsmath}
\usepackage{slashed}
\usepackage{amssymb}
\usepackage{graphicx}
\usepackage{makeidx}
\usepackage{cancel}
\usepackage{epic}
\usepackage{eepic}
\usepackage{epsfig}
\usepackage{latexsym}
\usepackage[dvipsnames]{xcolor}
\usepackage{float}
\usepackage{multirow}
\usepackage[export]{adjustbox}
\usepackage{xurl,hyperref}
\usepackage{enumitem}
\hypersetup{colorlinks=true,citecolor=red,linkcolor=NavyBlue,urlcolor=NavyBlue}
\usepackage[utf8]{inputenc}
\usepackage[caption=false]{subfig}
 
\usepackage{natbib}
\usepackage{relsize}
\usepackage[left=2.2 cm,right=2.2 cm,top=2.2 cm,bottom=2.2 cm]{geometry}
\usepackage{mathptmx}

\begin{document}
\relscale{1.05}
\captionsetup[subfigure]{labelformat=empty}

\title{A Compatibility Check: Low-Scale Chiral $U(1)_X$ Theories Vs. $(g-2)_e$ Anomaly}

\author{Bibhabasu De}
\email{bibhabasude@gmail.com}
\affiliation{Department of Physics, The ICFAI University Tripura, Kamalghat-799210, India}

\date{\today}

\begin{abstract}
\noindent
Chiral Abelian extensions of the Standard Model~(SM) gauge group may offer significant new possibilities for explaining various {\it Beyond the Standard Model}~(BSM) phenomena within a common framework. The models being less explored in the literature only a few experimental constraints have been reported to date, leaving a major portion of the parameter space available for the {\it New Physics}~(NP) phenomenology. The present paper considers three anomaly-free chiral Abelian extensions and examines the compatibility of the corresponding low-scale parameter spaces with the observed $(g-2)_e$ anomaly. The analysis results in stringent exclusion limits, completely ruling out the considered chiral models when used in complementarity with the existing experimental bounds.
\end{abstract}
	
\maketitle	

\section{Introduction}
\noindent
The Standard Model has proved itself as an extremely robust theoretical framework to explain the strong and electroweak~(EW) interactions. The precise experimental validation~\cite{ATLAS:2012yve, CMS:2012qbp,ATLAS:2024erm,CMS:2024lrd,LHCb:2025nob,Aliberti:2025beg} of the SM predictions has also played a significant role in consolidating the theory. However, the existence of a parallel BSM sector can't be ignored as the SM fails to address several observations, including dark matter~(DM)~\cite{1932BAN.....6..249O,Zwicky:1937zza,Metcalf:2003sz,Planck:2018vyg}, neutrino oscillations~\cite{Super-Kamiokande:1998kpq}, matter-antimatter asymmetry~\cite{Planck:2018vyg}, etc. Though there exists a plethora of BSM theories to explain these observations, a selectively/weakly interacting NP may lead to interesting phenomenological consequences and evade the current collider constraints even at the sub-GeV mass scales. Extending the SM gauge group $\mathcal{G}_{\rm SM}=SU(3)_C\otimes SU(2)_L\otimes U(1)_Y$ with an additional $U(1)_X$ symmetry is a well-motivated approach~\cite{Langacker:2008yv} to formulate such NP interactions. The associated physical gauge boson~($Z^\prime$) can mediate neutral current~(NC) interactions among the $U(1)_X$-charged fields, acting as a portal between the SM and BSM sectors. Such Abelian extensions naturally arise in the grand unified theories~\cite{Robinett:1982tq,Langacker:1984dc}, extra-dimensional models~\cite{Antoniadis:1990ew,Appelquist:2000nn}, string compactifications~\cite{Goodsell:2010ie}, or can be constituted by gauging some global symmetries of the classical SM Lagrangian. Depending on the transformation of the left- and right-chiral SM fermions under the considered $U(1)_X$ symmetry, the Abelian extensions can be labelled as either vector or chiral. The vector $U(1)_X$ extensions~\cite{Foot:1990mn,He:1991qd,Foot:1994vd,Mohapatra:1980de,Ma:1997nq,Appelquist:2002mw,Lee:2010hf,Bonilla:2017lsq,Alonso:2017uky} are well-known in the literature and have been extensively used to predict/describe various BSM observables~\cite{Fox:2008kb,Kopp:2009et, Essig:2011nj,Banerjee:2015hoa,Biswas:2016yan,Hasegawa:2019amx,Blanco:2019hah,De:2025hay,Okada:2010wd,Okada:2012fs,De:2024tvj,Ghosh:2024cxi,Ota:2006xr,Binetruy:1996cs,Baek:2001kca,Lam:2001fb,Choubey:2004hn,Wang:2019byi,Barman:2021yaz,Coloma:2020gfv,Farzan:2015doa,DeRomeri:2024dbv,Wise:2018rnb,Ardu:2022zom,Dasgupta:2023zrh,Asai:2019ciz,Borah:2021mri,Cai:2021evx,He:2024dwh,Koechler:2025ryv,De:2026hwq}. However, there is only a limited number of possibilities for the vector $U(1)_X$ models and the parameter spaces are shrinking continously. For example, two popular vector frameworks, namely $U(1)_{L_i-L_j}$~[$i,\,j=e,\,\mu,\,\tau$] and $U(1)_{B-L}$~[$B$ is the baryon number and $L=\sum\limits_{i} L_i$] can be strongly constrained through the ongoing and next-generation experiments, leading to substantial weakening of the NP interaction strength~[for recent reviews, see Refs.~\cite{Bauer:2018onh, Dasgupta:2023zrh}]. On the other hand, a quite fascinating but less explored framework is the chiral $U(1)_X$ extension of $\mathcal{G}_{\rm SM}$ where the left- and right-chiral fermions carry different $X$ charges~\cite{Prajapati:2024wuu}. For the chiral formulations, the anomaly cancellation~\cite{Adler:1969gk,Bardeen:1969md,Bell:1969ts,Delbourgo:1972xb,Alvarez-Gaume:1983ihn} can't be achieved trivially and one has to add extra BSM fermions to the particle spectrum. Moreover, to maintain the invariance of the SM Yukawa terms, the Higgs must transform non-trivially under the new chiral symmetry, resulting in a mass mixing between the EW and $U(1)_X$ gauge bosons. Therefore, the primary constraints to such extensions originate from the EW $\rho$ parameter~\cite{ParticleDataGroup:2024cfk} and the perturbativity. However, in general, the chiral $U(1)_X$ models can also be probed through various neutrino-based experiments, leading to stringent exclusion limits, specifically in the lighter mass regime. Thus, considering the current sensitivities of TEXONO~\cite{TEXONO:2009knm}, COHERENT~\cite{COHERENT:2021xmm,COHERENT:2020iec}, XENONnT~\cite{XENON:2022ltv}, LUX-ZEPLIN~(LZ)~\cite{LZ:2023poo}, PandaX-4T~\cite{PandaX:2024cic}, and the projected bounds from DARWIN~\cite{DARWIN:2020bnc},  Ref.~\cite{Majumdar:2024dms} has remarkably constrained the low-scale parameter spaces of three chiral $\mathcal{G}_{\rm SM}\otimes U(1)_X$ benchmark models~(BMs). At present, these bounds are the strongest ones for the considered chiral extensions.

The present paper picks up the same BMs~[will be discussed in Sec.~\ref{sec:chiral}] and checks if the allowed parameter spaces are compatible with the observed discrepancy between the measured and predicted values of the electron anomalous magnetic moment~($a_e=(g-2)_e/2$). Note that it is a generic feature of all the leptophilic Abelian extensions of $\mathcal{G}_{\rm SM}$ to induce loop-level corrections to the lepton anomalous magnetic moments through the gauge boson exchange. However, out of the three lepton $(g-2)$ values, at present, only the electron anomalous magnetic moment shows a significant discrepancy between its SM-predicted~\cite{Aoyama:2019ryr} and experimentally measured~\cite{Fan:2022eto} values. Though the earlier studies reported a similar discrepancy in the muon sector as well, the 2025 update~\cite{Muong-2:2025xyk,Aliberti:2025beg} has ruled out the possibility within the current level of experimental precision. Moreover, unlike the first two generations, measuring the tau anomalous magnetic moment is extremely challenging due to the short lifetime of tau. Thus, the existing bounds on $(g-2)_\tau$~\cite{DELPHI:2003nah} are effectively insignificant to constrain any NP theory. Therefore, considering the leading-order BSM contribution to $(g-2)_e$, the present work shows that almost the entire low-scale parameter space of the chiral $U(1)_X$ theories can be excluded. Further, the considered BMs can be completely ruled out if the detector constraints~\cite{Majumdar:2024dms} are also incorporated in the analysis.

The rest of the paper has been structured as follows. Sec.~\ref{sec:chiral} discusses the general anomaly cancellation conditions for a chiral $\mathcal{G}_{\rm SM}\otimes U(1)_X$ framework leading to the aforementioned BMs, formulates the physical gauge eigenstates and the NP interactions among the SM leptons. The resultant one-loop BSM corrections to the electron anomalous magnetic moment have been explored in Sec.~\ref{sec:amu}, followed by Sec.~\ref{sec:ex} where the BMs have been constrained through the $(g-2)_e$ anomaly, and the corresponding exclusion limits have been plotted against the existing experimental bounds for a comparison. Finally, the work has been concluded in Sec.~\ref{sec:conc}.     
\section{Building a Chiral $U(1)_X$ Model}
\noindent
\label{sec:chiral}
The paper extends the SM gauge group with a chiral Abelian symmetry, $U(1)_X$. The new symmetry is chiral in the sense that the left- and right-handed SM fermions possess different $X$ charges, i.e., they are in different representations of the $U(1)_X$ gauge group. However, such BSM extensions introduce gauge anomalies in the theory, and one has to rigorously solve the anomaly cancellation conditions to fix the $X$ charges of all the chiral fermions. Though for the hypercharge group, anomaly cancellation can be achieved with the SM fermions alone, for $\mathcal{G}_{\rm SM}\otimes U(1)_X$, three additional right-handed SM-singlet fermions, with non-trivial $X$ charges, are required. Let's call them $\psi_k$~[$k=1,\,2,\,3$] with $X_\psi^{(k)}$ being their $U(1)_X$ charges, respectively. Thus, following Ref.~\cite{Prajapati:2024wuu}, the triangle and mixed gauge-gravitational anomaly cancellation conditions for the proposed framework can be cast as,
\begin{align}
&[SU(3)_C]^2[U(1)_X]\Rightarrow \sum_{i\,=1}^3\Big[2X_Q^{(i)}-X_{u_R}^{(i)}-X_{d_R}^{(i)}\Big]=0\nonumber\\
&[SU(2)_L]^2[U(1)_X]\Rightarrow \sum_{i\,=1}^3\Big[X_\mathbb{L}^{(i)}+3X_Q^{(i)}\Big]=0\nonumber\\
&[U(1)_Y]^2[U(1)_X]\Rightarrow \sum_{i\,=1}^3\Big[3X_\mathbb{L}^{(i)}+X_Q^{(i)}-6X_{\ell_R}^{(i)}-8X_{u_R}^{(i)}-2X_{d_R}^{(i)}\Big]=0\nonumber\\
&[U(1)_Y][U(1)_X]^2\Rightarrow \sum_{i\,=1}^3\left[\left\{X_Q^{(i)}\right\}^2-\left\{X_\mathbb{L}^{(i)}\right\}^2+\left\{X_{\ell_R}^{(i)}\right\}^2-2\left\{X_{u_R}^{(i)}\right\}^2+\left\{X_{d_R}^{(i)}\right\}^2\right]=0\nonumber\\
&[U(1)_X]^3\Rightarrow \sum_{i\,=1}^3\left[2\left\{X_\mathbb{L}^{(i)}\right\}^3+6\left\{X_Q^{(i)}\right\}^3-\left\{X_{\ell_R}^{(i)}\right\}^3-3\left(\left\{X_{u_R}^{(i)}\right\}^3+\left\{X_{d_R}^{(i)}\right\}^3\right)\right]-\sum_{k\,=1}^3\left\{X_\psi^{(k)}\right\}^3=0\nonumber\\
&[{\rm Gravity}]^2[U(1)_X]\Rightarrow \sum_{i\,=1}^3\left[2X_\mathbb{L}^{(i)}+6X_Q^{(i)}-X_{\ell_R}^{(i)}-3\left(X_{u_R}^{(i)}+X_{d_R}^{(i)}\right)\right]-\sum_{k\,=1}^3X_\psi^{(k)}=0\,.
\label{eq:anom}
\end{align}
Here $i$ stands for the SM flavor index. $X_{\ell_R}$, $X_{u_R}$, and $X_{d_R}$ denote the $X$ charges for right-handed SM leptons, up-type and down-type quarks, respectively, while $X_\mathbb{L}$ and $X_Q$ are the corresponding charges for left-handed lepton and quark doublets. Note that in addition to Eq.~\eqref{eq:anom}, the invariance of Yukawa terms plays a vital role in defining the $U(1)_X$ charges of the chiral fermions and the SM Higgs~($H$). Thus, the SM fermions can acquire masses through the electroweak symmetry breaking~(EWSB) if and only if
\begin{align}
X_H=X_{\mathbb{L}}-X_{\ell_R}=X_{u_R}-X_Q=X_Q-X_{d_R}\,,
\label{eq:XH}
\end{align}
where $X_H$ is the new Abelian charge of $H$. Combining Eqs.~\eqref{eq:anom} and \eqref{eq:XH} for a flavor-blind scenario~(in the SM sector), one can obtain the following relations.
\begin{align}
&X_Q=-\frac{X_\mathbb{L}}{3}\,,\qquad X_{u_R}=\frac{2X_\mathbb{L}}{3}-X_{\ell_R}\,,\qquad X_{d_R}=-\frac{4X_\mathbb{L}}{3}+X_{\ell_R}\,,\qquad X_H=X_{\mathbb{L}}-X_{\ell_R}\,,\nonumber\\
&\sum_{k\,=1}^3\left\{X_\psi^{(k)}\right\}^3=3\left(2X_\mathbb{L} - X_{\ell_R}\right)^3\,,\qquad \sum_{k\,=1}^3X_\psi^{(k)}=3\left(2X_\mathbb{L} - X_{\ell_R}\right)\,.
\label{eq:charges}
\end{align}
Table~\ref{tab:sol} enlists four distinct classes of solutions for the above equations, where each class can result in a set of chiral $U(1)_X$ models for different values of the chosen free parameters\,\footnote{The solutions were originally proposed in Ref.~\cite{Prajapati:2024wuu}.}. These four classes will be referred to as type-I, II, III, and IV, resepectively.
\begin{table}[!ht]
\centering
\begin{tabular}{|c|c|c|c|c|c|c|c|c|c|}
\hline
Solution & $X_Q$ & $X_{u_R}$ & $X_{d_R}$ & $X_{\mathbb{L}}$ & $X_{\ell_R}$ & $X_\psi^{(1)}$ & $X_\psi^{(2)}$ & $X_\psi^{(3)}$ & $X_H$\\ 
Type & & & & & & & & & \\
\hline\hline
I & $-\frac{X_\mathbb{L}}{3}$ & $-\frac{4X_\mathbb{L}}{3}$ & $\frac{2X_\mathbb{L}}{3}$ & $~X_\mathbb{L}~$ & $2X_\mathbb{L}$ & $0$ & $\zeta$ & $-\zeta$ & $-X_\mathbb{L}$\\
\hline
II & $-\frac{X_\mathbb{L}}{3}$ & $-\frac{4X_\mathbb{L}}{3}+\zeta$ & $\frac{2X_\mathbb{L}}{3}-\zeta$ & $~X_\mathbb{L}~$ & $2X_\mathbb{L}-\zeta$ & $\zeta$ & $\zeta$ & $\zeta$ & $\zeta-X_\mathbb{L}$\\
\hline
III & $\frac{1}{\eta}$ & $-\,\frac{(\eta\zeta-4)}{\eta}$ & $\frac{\eta\zeta-2}{\eta}$ & $-\frac{3}{\eta}$ & $\frac{\eta\zeta-6}{\eta}$ & $5\zeta$ & $-4\zeta$ & $-4\zeta$ & $-\frac{(\eta\zeta-3)}{\eta}$\\
\hline
IV & $-\frac{X_\mathbb{L}}{3}$ & $-\frac{4X_\mathbb{L}}{3}$ & $\frac{2X_\mathbb{L}}{3}$ & $~X_\mathbb{L}~$ & $2X_\mathbb{L}$ & $~\frac{5\eta^2+3\zeta^2}{8}~$ & $-\frac{\eta^2}{2}$ & $-\frac{\eta^2}{2}$ & $-X_\mathbb{L}$\\
 & & $-\left(\frac{\eta^2-\zeta^2}{8}\right)$ & $+\frac{\eta^2-\zeta^2}{8}$ & & $+\frac{\eta^2-\zeta^2}{8}$ & & $-\left(\frac{3\eta^2\zeta+\zeta^3}{8\eta}\right)$ & $+\left(\frac{3\eta^2\zeta+\zeta^3}{8\eta}\right)$ & $-\left(\frac{\eta^2-\zeta^2}{8}\right)$ \\
\hline
\end{tabular}
\caption{$U(1)_X$ charges of all the fermions and SM Higgs, consistent with the anomaly cancellation conditions and the SM Yukawa interactions. $X_{\mathbb{L}}$, $\eta$, and $\zeta$ are the free parameters and can be varied to obtain different chiral $U(1)_X$ models.}
\label{tab:sol}
\end{table}
However, for the numerical analysis, particular BMs are required. Table~\ref{tab:BM} represents three anomaly-free chiral $U(1)_X$ models, first introduced in Ref.~\cite{Majumdar:2024dms}, where the SM Higgs is sufficient to generate the masses of all the SM fermions. 
\begin{table}[!ht]
\centering
\begin{tabular}{|c|c|c|c|c|c|c|c|c|c|}
\hline
Benchmark & $~X_Q~$ & $~X_{u_R}~$ & $~X_{d_R}~$ & $~X_{\mathbb{L}}~$ & $~X_{\ell_R}~$ & $~X_\psi^{(1)}~$ & $~X_\psi^{(2)}~$ & $~X_\psi^{(3)}~$ & $~X_H~$\\ 
Models & & & & & & & & & \\
\hline\hline
{\bf BM1} & $-\,13/3$ & $-\,4/3$ & $-\,22/3$ & $13$ & $10$ & $16$ & $16$ & $16$ & $3$\\
\hline
{\bf BM2} & $-\,3$ & $10$ & $-\,16$ & $9$ & $-\,4$ & $-\,110$ & $88$ & $88$ & $13$\\
\hline
{\bf BM3} & $-\,1/3$ & $5/3$ & $-\,7/3$ & $1$ & $-\,1$ & $10$ & $-\,18$ & $17$ & $2$\\
\hline
\end{tabular}
\caption{Anomaly-free chiral $U(1)_X$ models. BM1, BM2, and BM3 correspond to type-II, III, and IV solutions, respectively.}
\label{tab:BM}
\end{table}
To be specific, if one chooses $X_\mathbb{L}=1$, $\eta=1$, and $\zeta=5$ in the type-IV solution of Table~\ref{tab:sol}, BM3 can be obtained. Similarly, BM1 corresponds to $X_\mathbb{L}=13$ and $\zeta=16$ in the type-II solution, while BM2 can be mapped into the type-III for $\eta=-1/3$ and $\zeta=-22$. Type-I leads to the hypercharge-like solutions where the anomalies from the BSM fermions get mutually canceled. Note that the additional BSM fermions significantly enrich the phenomenology of the considered framework. For example, the lightest of the three SM-singlet fermions can serve as a viable DM candidate~\cite{Prajapati:2024wuu}. The chiral $U(1)_X$ models also have the potential to explain neutrino osciallaions~\cite{Prajapati:2026tfv}, $B$-anomalies~\cite{Prajapati:2026tfv}, and ATOMKI anomaly~\cite{Batra:2026tzz}. However, these discussions are beyond the scope of the present work and might be explored later.
\subsection{Gauge Sector}
In the case of chiral theories, $H$ being unavoidably charged under $U(1)_X$, there is a mixing between the EW gauge bosons and the new Abelian gauge boson, $\mathbb{C}$, in the flavor basis. Thus, the covariant derivative for a chiral $\mathcal{G}_{\rm SM}\otimes U(1)_X$ theory can be formulated as,
\begin{align}
\mathcal{D}_\alpha=\partial_\alpha+ig_3\frac{\Lambda^j}{2}G^j_\alpha+ig_2\frac{\sigma^b}{2}W^b_\alpha+ig_1\frac{Y}{2}B_\alpha+ig_XX\mathbb{C}_\alpha\,,
\end{align}
where $g_i$~[$i=1,\,2,\,3$] are the gauge couplings associated with $SU(3)_C$, $SU(2)_L$, and $U(1)_Y$, respectively while $g_X$ is the gauge coupling for $U(1)_X$ gauge group. $\Lambda^j$~[$j=1,\,\cdots,\, 8$] and $\sigma^b$~[$b=1,\,2,\,3$] denote the Gell-Mann and Pauli matrices, respectively. Therefore, the scalar sector can be defined as,
\begin{align}
\mathcal{L}_{\rm Scalar}=(\mathcal{D}^\alpha H)^\dagger (\mathcal{D}_\alpha H)+(\mathcal{D}^\alpha \Phi)^\dagger (\mathcal{D}_\alpha \Phi)-V(\Phi,\, H)\,.
\end{align}
Here $\Phi$ is an SM-singlet scalar, non-trivially charged under the $U(1)_X$ symmetry. Note that the presence of this extra scalar field is necessary for keeping the $U(1)_X$ and EW symmetry breaking scales independent of each other. Moreover, with an appropriately assigned charge, $\Phi$ can generate masses of the BSM fermions once the $U(1)_X$ symmetry is spontaneously broken. $V(\Phi,\, H)$ is the usual scalar potential for a theory with ``{\it $H+$ complex SM-singlet scalar}"-like structure~\cite{Barger:2008jx}. After spontaneous symmetry breaking~(SSB), the scalar fields can be redefined as,
\begin{align}
H=\frac{1}{\sqrt{2}}\left(\begin{array}{c}
0\\
h+v
\end{array}\right),\qquad\qquad \Phi=\frac{(\phi+u)}{\sqrt{2}}\,,
\end{align}
where $h$ and $\phi$ are the physical scalar fields with masses $M_h= 125$ GeV and $M_\phi$, while $v= 246.22$ GeV and $u$ represent the EW and $U(1)_X$ vacuum expectation values~(VEVs), respectively. Thus, after SSB, the gauge boson mass matrix in the $\left(B_\alpha,\,W^3_\alpha,\,\mathbb{C}_\alpha\right)$ basis can be cast as,
\begin{align}
\mathbb{M}^2=\frac{v^2}{4}\left[\begin{array}{c c c}
g_1^2 & -g_1g_2 & 2g_1g_XX_H\\
-g_1g_2 & g_2^2 & -2g_2g_XX_H\\
2g_1g_XX_H & -2g_2g_XX_H & 4g_X^2\lambda^2
\end{array}\right]\,,
\label{eq:mass}
\end{align}
where $\lambda=\sqrt{X_H^2+X_\Phi^2\left(u^2/v^2\right)}$ with $X_\Phi$ being the $U(1)_X$ charge of $\Phi$. Note that for simplicity, kinetic mixing has been ignored in the present work. The physical basis $\left(A_\alpha,\,Z_\alpha,\, (Z^\prime)_\alpha\right)$ can be obtained via a 3-dimensional orthogonal rotation as follows~\cite{Majumdar:2024dms,Prajapati:2026tfv}.
\begin{align}
\left(\begin{array}{c}
A_\alpha\\
Z_\alpha\\
(Z^\prime)_\alpha
\end{array}\right)=\left[\begin{array}{c c c}
\cos \theta_W & \sin\theta_W & 0\\
-\cos\theta_X\sin\theta_W & \cos\theta_X\cos\theta_W & -\sin\theta_X\\
-\sin\theta_X\sin\theta_W & \sin\theta_X\cos\theta_W & \cos\theta_X
\end{array}\right]\left(\begin{array}{c}
B_\alpha\\
W^3_\alpha\\
\mathbb{C}_\alpha
\end{array}\right)\,.
\label{eq:rotn}
\end{align}
Here $\theta_W=\tan^{-1}(g_1/g_2)$ is the weak mixing angle and $\theta_X$ parametrizes the mixing between $Z$ and $Z^\prime$. After diagonalization, the mass eigenvalues are given by
\begin{align}
M_A^2=0,\qquad M_Z^2=\frac{v^2}{8}\left(\Omega_1+\sqrt{\Omega_2^2+\Omega_3^2}\right),\qquad M_{Z^\prime}^2=\frac{v^2}{8}\left(\Omega_1-\sqrt{\Omega_2^2+\Omega_3^2}\right)\,.
\end{align}
The massless gauge boson $A$ can be identified as the photon. The parameters $\Omega_i$~[$i=1,\,2,\,3$] stand for,
\begin{align}
\Omega_1=4\left[\left(\frac{M_Z^{\rm SM}}{v}\right)^2+g_X^2\lambda^2\right],\quad \Omega_2=8g_XX_H\left(\frac{M_Z^{\rm SM}}{v}\right),\quad\Omega_3=4\left[\left(\frac{M_Z^{\rm SM}}{v}\right)^2-g_X^2\lambda^2\right]\,,
\end{align}
where $M_Z^{\rm SM}=\lim\limits_{g_X\to 0}\,M_Z=\frac{v}{2}\left(g_1^2+g_2^2\right)^{1/2}$ represents the SM value of the $Z$ boson mass. Note that the $U(1)_X$ symmetry-breaking scale $u$ and the gauge boson mass $M_{Z^\prime}$ are correlated. Thus, considering $M_{Z^\prime}$ a free parameter, one can eliminate the redundant mass-dimensional quantity $X_\Phi u$ and recast $\lambda$ as,
\begin{align}
\lambda=\left[X_H^2+\frac{M^2_{Z^\prime}}{g_X^2v^2}\left\{\frac{\left(M_Z^{\rm SM}\right)^2+g_X^2X_H^2v^2-M_{Z^\prime}^2}{\left(M_Z^{\rm SM}\right)^2-M_{Z^\prime}^2}\right\}\right]^{1/2}\,.
\end{align}
In the present paper, $M_Z$ has been considered as the heavier state, i.e., all the ensuing analysis will correspond to a parameter space where $M_{Z^\prime}<M_Z$. Moreover, it is trivial to check that $\lim\limits_{g_X\to 0}\,M_{Z^\prime}=0$. Finally, the mixing angle $\theta_X$ can be defined as,
\begin{align}
\theta_X=\frac{1}{2}\sin^{-1}\left[\frac{2vg_XX_H\,M_Z^{\rm SM}}{M_Z^2-M_{Z^\prime}^2}\right]\,.
\label{eq:theta}
\end{align}
However, this mass mixing can't be arbitrarily large. For any non-zero value of $\theta_X$, the $\rho$ parameter deviates from its SM-predicted value, which closely agrees with the experimental measurements. The $\rho$ parameter is defined as, $\rho=M_W^2/(M_Z^2\cos^2\theta_W)$ with $\rho_{\rm SM}=1$ for tree-level interactions. In the considered gauge extension, the $W$ boson mass does not deviate from its SM value, whereas the $Z$ boson mass increases due to mixing. Therefore, the $\rho$ parameter can be expressed as~\cite{Altarelli:1991zg,Bento:2023weq},
\begin{align}
\rho=1+\left[\left(\frac{M_{Z^\prime}}{M_Z}\right)^2-1\right]\sin^2\theta_X\,.
\end{align}
\begin{figure}[!ht]
\centering
\subfloat[\qquad\qquad(a)]{\includegraphics[scale=0.65]{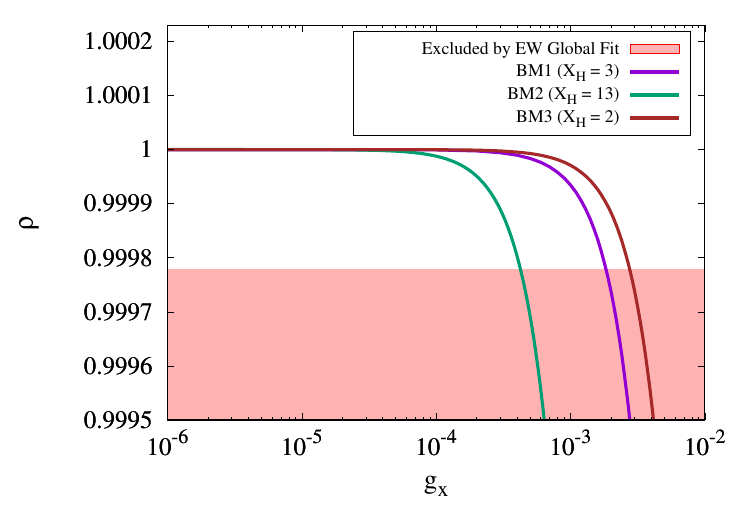}}
\subfloat[\qquad\qquad(b)]{\includegraphics[scale=0.65]{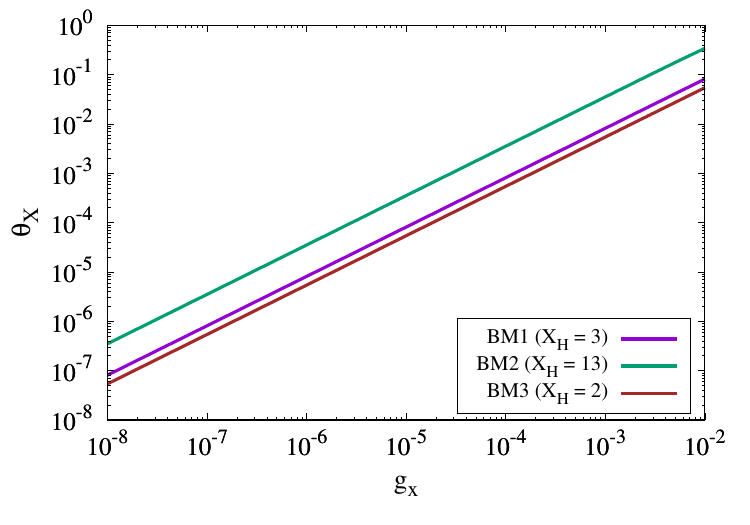}}
\caption{Variation of (a) $\rho$ and (b) $\theta_X$ as a function of $g_X$ for $M_{Z^\prime}=1$ GeV. The colors violet, green, and brown represent the BM1, BM2, and BM3, respectively.}
\label{fig:rho}
\end{figure}
Fig.~\ref{fig:rho}\,(a) shows the variation of $\rho$ parameter as a function of $g_X$ with the $M_{Z^\prime}$ being fixed at 1 GeV. The three colors violet, green, and brown symbolize $X_H=3$~(BM1), $X_H=13$~(BM2), and $X_H=2$~(BM3), respectively. The global fit to the EW precision data results in $\rho=1.00038 \pm 0.00060\,(3\sigma)$~\cite{ParticleDataGroup:2024cfk}. Thus, the red shade in Fig.~\ref{fig:rho}\,(a) denotes the region with $\rho<0.99978$. Fig.~\ref{fig:rho}\,(b) depicts the variation of $\theta_X$ with the $U(1)_X$ gauge coupling for $M_{Z^\prime}=1$ GeV. The graphical behaviour can be easily explained from Eq.~\eqref{eq:theta}. However, for $g_X>10^{-2}$, $M_Z$ starts to rise sharply resulting in a significant deviation from $M_Z^{\rm SM}$. For such strongly interacting chiral theories, the mixing angle $\theta_X$ shows a decreasing trend as one goes beyond $g_X=10^{-2}$. Nonetheless, the parameter space with $g_X>10^{-2}$ is phenomenologically irrelevant as it corresponds to a violation of the $\rho$ parameter constraint for all three BMs. Note that both $\theta_X$ and $\rho$ do not exhibit a significant mass dependence for the low-scale chiral theories. However, in the heavier mass regime where $M_{Z^\prime}>M_Z$, both of the parameters fall as $M_{Z^\prime}$ increases, asymptotically tending to zero and $\rho_{\rm SM}$, respectively~\cite{Prajapati:2024wuu}.

\subsection{Interaction with the SM Leptons}
As the primary goal of the present paper is to constrain the considered chiral BMs using the $(g-2)_e$ anomaly, it is crucial to formulate the interactions of $Z$ and $Z^\prime$ with the SM leptons. From Eq.~\eqref{eq:rotn}, one can obtain,
\begin{align}
B_\alpha=&~A_\alpha\cos\theta_W-Z_\alpha\cos\theta_X\sin\theta_W-Z^\prime_\alpha\sin\theta_X\sin\theta_W\,,\nonumber\\
W^3_\alpha=&~A_\alpha\sin\theta_W+Z_\alpha\cos\theta_X\cos\theta_W+Z^\prime_\alpha\sin\theta_X\cos\theta_W\,,\nonumber\\
\mathbb{C}_\alpha=&~-Z_\alpha\sin\theta_X+Z^\prime_\alpha\cos\theta_X\,.
\label{eq:phybas}
\end{align}
Further, in the present gauge configuration, the NC and electromagnetic~(EM) interactions of the SM leptons stem from,
\begin{align}
-\mathcal{L}_{\rm NC+EM}=&~\bar{\mathbf{L}}\gamma^\alpha\left[g_2T_3W^3_{\alpha}+\frac{g_1Y}{2}B_\alpha+g_XX\mathbb{C}_\alpha\right]\mathbf{L}\,,
\label{eq:NC}
\end{align}
where $T_3$ is the third component of the weak isospin, while the EM charge can be defined as $Q=T_3+Y/2$. The notation $\mathbf{L}$ stands for any generic SM lepton field such that $P_L\mathbf{L}\equiv \mathbb{L}=(\nu\quad\ell)_L^T$ and $P_R\mathbf{L}\equiv\ell_R$ define the left-handed lepton doublets and right-handed lepton singlets, respectively. $P_{L,\,R}$ represent the chiral projection operators. Combining Eqs.~\eqref{eq:phybas} and \eqref{eq:NC}, the aforesaid interactions can be expressed in the physical basis as,
\begin{align}
-\mathcal{L}_{\rm NC+EM}=&~\bar{\mathbf{L}}\gamma^\alpha\Bigg[A_\alpha\Big(g_2T_3\sin\theta_W+\frac{g_1Y}{2}\cos\theta_W\Big)+Z_\alpha\Big\{-g_XX\sin\theta_X+\Big(g_2T_3\cos\theta_W\nonumber\\
&-\frac{g_1Y}{2}\sin\theta_W\Big)\cos\theta_X\Big\}+Z^\prime_\alpha\Big\{g_XX\cos\theta_X+\Big(g_2T_3\cos\theta_W-\frac{g_1Y}{2}\sin\theta_W\Big)\sin\theta_X\Big\}\Bigg]\mathbf{L}\nonumber\\
=&~\bar{\mathbf{L}}\left[eQ_{\mathbf{L}}\gamma^\alpha\, A_\alpha\right]\mathbf{L}+\bar{\mathbf{L}}\left[\gamma^\alpha\left(\mathbb{S}_L^ZP_L+\mathbb{S}_R^ZP_R\right)\, Z_\alpha\right]\mathbf{L}+\bar{\mathbf{L}}\left[\gamma^\alpha\left(\mathbb{S}_L^{Z^\prime}P_L+\mathbb{S}_R^{Z^\prime} P_R\right)\, Z^\prime_\alpha\right]\mathbf{L}\,,
\label{eq:lep}
\end{align}
where the first term represents the usual EM interactions with $Q_\mathbf{L}=-1$ for the charged leptons and $0$ for the neutrinos. The chiral couplings in the second and third terms can be defined as,
\begin{align}
\mathbb{S}_L^Z=&~\frac{2\,M_Z^{\rm SM}}{v}\left(T_3-Q_\mathbf{L}\sin^2\theta_W\right)\cos\theta_X-X_{\mathbb{L}}g_X\sin\theta_X\,,\nonumber\\
\mathbb{S}_R^Z=&~-\frac{2\,M_Z^{\rm SM}}{v}\times Q_\mathbf{L}\sin^2\theta_W\cos\theta_X-X_{\ell_R}g_X\sin\theta_X\,,\nonumber\\
\mathbb{S}_L^{Z^\prime}=&~\frac{2\,M_Z^{\rm SM}}{v}\left(T_3-Q_\mathbf{L}\sin^2\theta_W\right)\sin\theta_X+X_{\mathbb{L}}g_X\cos\theta_X\,,\nonumber\\
\mathbb{S}_R^{Z^\prime}=&~-\frac{2\,M_Z^{\rm SM}}{v}\times Q_\mathbf{L}\sin^2\theta_W\sin\theta_X+ X_{\ell_R}g_X\cos\theta_X\,.
\end{align}
Note that in the above expressions, $g_2/\cos\theta_W$ has been recast as $(2\,M_Z^{\rm SM})/v$. 
\section{Correction to $(g-2)_e$}
\noindent
\label{sec:amu}
The electron anomalous magnetic moment is experimentally well measured and, at present, can serve as a vital constraint on leptophilic NP theories. The latest experimental update for $a_e$ can be read as $a_e^{\rm Exp}=1 159 652 180.59(13)\times 10^{-12}$~\cite{Fan:2022eto}. However, the best available SM predictions for $a_e$ follow from the data-driven perturbative methods that rely upon the measurement of the fine-structure constant using the recoil of the atoms through photon absorption~\cite{Aoyama:2019ryr}. Presently, the measurements corresponding to Rubidium-87~\cite{Morel:2020dww} and Cesium-133~\cite{Parker:2018vye} show a $5.5\sigma$ discrepancy between the resultant values of $a_e^{\rm SM}$. Thus, the corresponding deviations from the experiment can be defined as,
\begin{align}
\Delta \,a_e^{\rm Cs}&=(-8.8\pm 3.6)\times 10^{-13}~(2.4\sigma)\,,\nonumber\\
\Delta \,a_e^{\rm Rb}&=(4.8\pm 3.0)\times 10^{-13}~(1.6\sigma)\,.
\end{align} 
The ambiguity between the two reported values of $a_e^{\rm SM}$ might only be settled through future experiments. Note that, though the lattice-QCD predictions for $a_e^{\rm SM}$ are also available in the literature~\cite{Budapest-Marseille-Wuppertal:2017okr,Giusti:2019hkz,Giusti:2020efo}, their precision is not yet comparable to that of data-driven computations.  
\begin{figure}[!ht]
\centering
\includegraphics[scale=0.6]{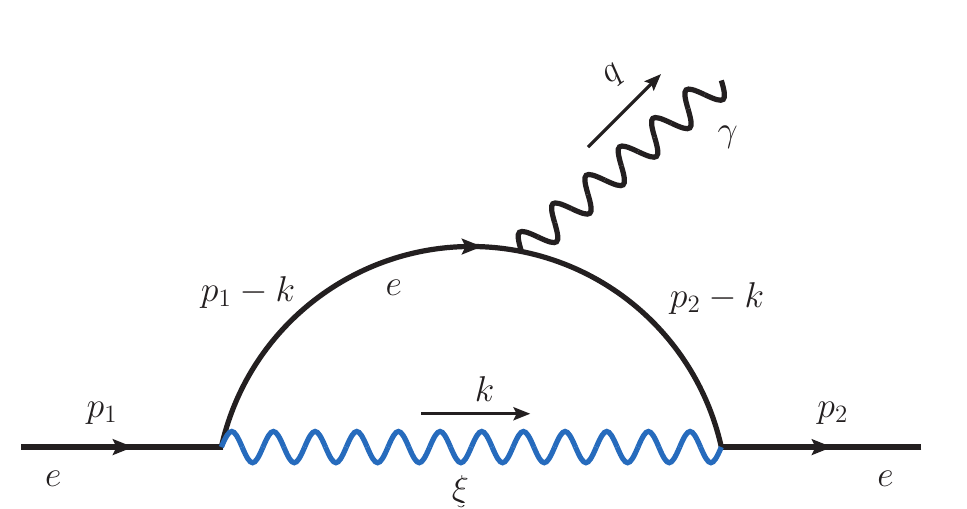}
\caption{One-loop correction to the $\bar{e}e\gamma$ vertex in the presence of a neutral massive gauge boson $\xi$. $q=p_1-p_2$ is the momentum of the on-shell photon.}
\label{fig:loop}
\end{figure}

The vertex correction factor corresponding to Fig.~\ref{fig:loop} can be defined as,
\begin{align}
\bar{u}(p_2)\left(-ie\,\delta\Gamma^\mu_\xi\right)&u(p_1)=-ie\int\frac{d^4k}{(2\pi)^4}\,\bar{u}(p_2)\Bigg[i\left(\left[\mathcal{B}^\xi_V\right]^*\gamma^\alpha+\left[\mathcal{B}^\xi_A\right]^*\gamma^\alpha\gamma_5\right)\,\frac{i(\slashed{p}_2-\slashed{k}+m_e)}{(p_2-k)^2-m_e^2}~\gamma^\mu\nonumber\\
&\times\frac{i(\slashed{p}_1-\slashed{k}+m_e)}{(p_1-k)^2-m_e^2}\,i\left(\mathcal{B}^\xi_V\gamma^\beta+\mathcal{B}^\xi_A\gamma^\beta\gamma_5\right)\,\frac{-i}{k^2-M_\xi^2}\left(g_{\alpha\beta}-\frac{k_\alpha k_\beta}{M_\xi^2}\right)\Bigg]u(p_1)\,.
\label{eq:vert}
\end{align}
Here $\xi$ stands for a generic neutral massive gauge boson with chiral couplings to the SM fermions, and, $\mathcal{B}_V^\xi=\left(\mathbb{S}_R^\xi+\mathbb{S}_L^\xi\right)/2$ and $\mathcal{B}_A^\xi=\left(\mathbb{S}_R^\xi-\mathbb{S}_L^\xi\right)/2$ represent the corresponding vector and axial-vector couplings, respectively. In the considered framework, $\xi=Z,\,Z^\prime$. After Feynman parametrization, Eq.~\eqref{eq:vert} can be recast as,
\begin{align}
\bar{u}(p_2)\left(\delta\Gamma^\mu_\xi\right)&u(p_1)=-2i\int^1_0 dx\int^{1-x}_0 dy\int\frac{d^4\ell}{(2\pi)^4}\,\bar{u}(p_2)\Bigg[\frac{\mathbb{N}^\mu(x,y,\ell)}{(\ell^2-\Delta_\xi)^3}\Bigg]u(p_1)\,,
\end{align}
where $\Delta_\xi=M_\xi^2[x+(1-x)^2R_{e\xi}]$, with $R_{e\xi}=(m_e/M_\xi)^2$. In general, the numerator $\mathbb{N}^\mu(x,y,\ell)$ contains a large number of terms. However, for the NP scale $M_\xi\geq 10^{-2}$ GeV, $R_{e\xi}$ is negligible compared to unity, leading to a significant simplification of the numerator. Thus, at the $\mathcal{O}(R_{e\xi})$, the one-loop contribution to $(g-2)_e$ can be approximated as~\cite{Yu:2021suw},
\begin{align}
a_{e\xi}=\frac{R_{e\xi}}{12\pi^2}\Bigg(\left|\mathcal{B}^\xi_V\right|^2-5\left|\mathcal{B}^\xi_A\right|^2\Bigg)\,.
\label{eq:g2mu1}
\end{align}
Note that in the chiral $U(1)_X$ extensions, there are two BSM contributions to the electron anomalous magnetic moment corresponding to the $Z$ and $Z^\prime$-induced one-loop diagrams. However, both of these gauge bosons contain some fraction of $Z_{\rm SM}$. Thus, one has to subtract the $Z_{\rm SM}$-contribution to $(g-2)_e$\,\footnote{The $Z_{\rm SM}$-contribution is already present in the definition of $a_e^{\rm SM}$. Therefore, this subtraction is necessary to avoid double counting.} from $a_{e Z}+a_{e Z^\prime}$ to obtain the actual BSM correction to the electron anomalous magnetic moment. Therefore, 
\begin{align}
\Delta\, a_e=a_{e Z}+a_{e Z^\prime}-a_{e Z_{\rm SM}}\,.
\end{align}  
\begin{figure}[!ht]
\centering
\subfloat[\qquad\qquad(a)]{\includegraphics[scale=0.65]{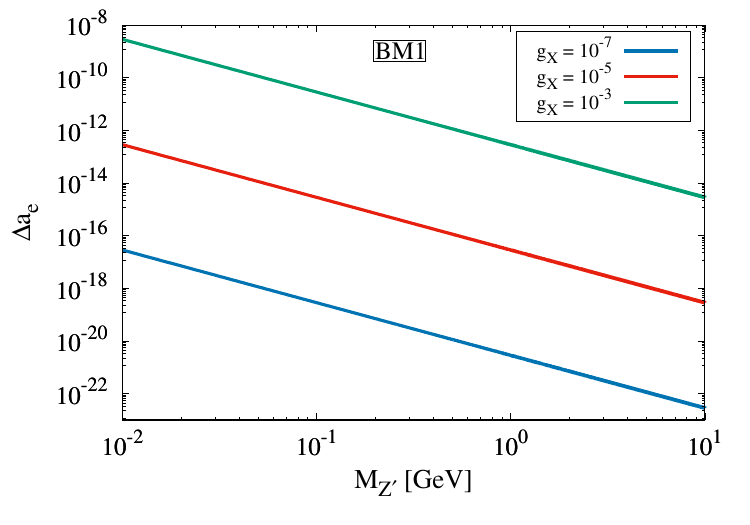}}
\subfloat[\qquad\qquad(b)]{\includegraphics[scale=0.65]{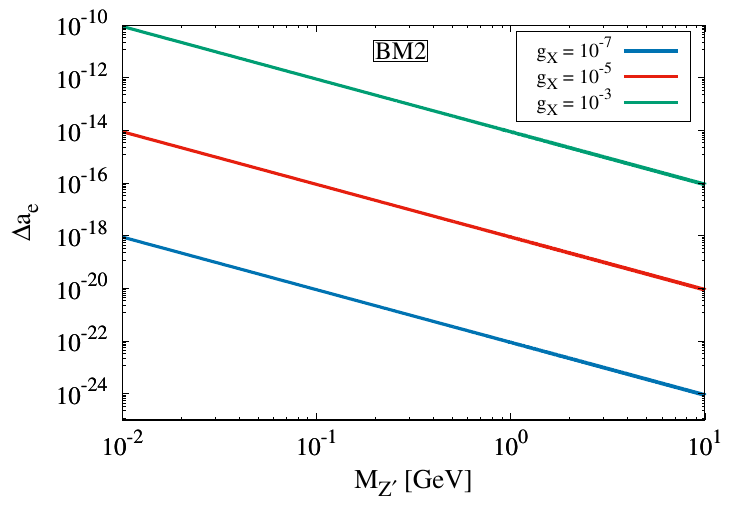}}\\
\subfloat[\qquad\qquad(c)]{\includegraphics[scale=0.65]{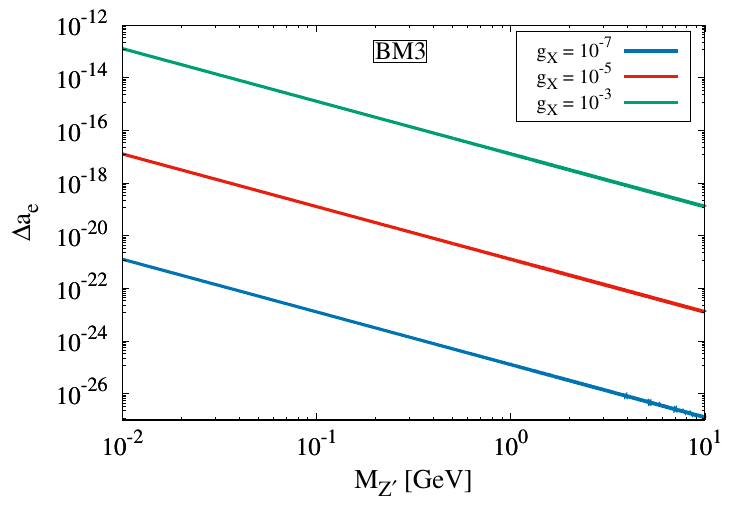}}
\caption{Variation of $\Delta\, a_e$ as a function of $M_{Z^\prime}$ for (a) BM1~($X_{e_L}=13$, $X_{e_R}=10$), (b) BM2~($X_{e_L}=9$, $X_{e_R}=-4$), and (c) BM3~($X_{e_L}=1$, $X_{e_R}=-1$). The three different colors represent $g_X=10^{-7}$~(blue), $g_X= 10^{-5}$~(red), and $g_X=10^{-3}$~(green).}
\label{fig:DA_M}
\end{figure}
Fig.~\ref{fig:DA_M} depicts the variation of $\Delta\, a_e$ as a function of $M_{Z^\prime}$ for the three BMs. The colors blue, red, and green represent $g_X=10^{-7}$, $g_X= 10^{-5}$, and $g_X=10^{-3}$, respectively. Note that the major BSM contribution to $\Delta \,a_e$ originates from the $Z^\prime$-exchange diagram, as due to a large cancellation between $a_{e Z}$ and $a_{e Z_{\rm SM}}$, the $Z$ boson contribution becomes subdominant. It has been numerically verified that $a_{e Z}-a_{e Z_{\rm SM}}$ is negative and approximately $\mathcal{O}(10^4)$ suppressed compared to $a_{e Z^\prime}$. Moreover, for the considered models, $\left|\mathcal{B}^{Z^\prime}_V\right|>\sqrt{5}\left|\mathcal{B}^{Z^\prime}_A\right|$ over the entire range of parameters, resulting in a positive correction to the electron anomalous magnetic moment. Thus, $\Delta\, a_e^{\rm Rb}$ can only be used to derive meaningful constraints on the parameter space. From Eq.~\eqref{eq:g2mu1}, as well as from the plots, it is evident that the theory shows a perfect decoupling nature as $\Delta\, a_e$ falls with the NP scale. Further, with decreasing $X_\mathbb{L}$ and/or $X_{\ell_R}$ values, the magnitude of $\Delta\,a_e$ drops for a given $\left\{M_{Z^\prime},\,g_X\right\}$ point. The trend stems from the coupling factors $\mathcal{B}_V^{Z^\prime}$ and $\mathcal{B}_A^{Z^\prime}$ which significantly depend upon the $U(1)_X$ charges of the considered SM lepton. 
\begin{figure}[!ht]
\centering
\includegraphics[scale=0.65]{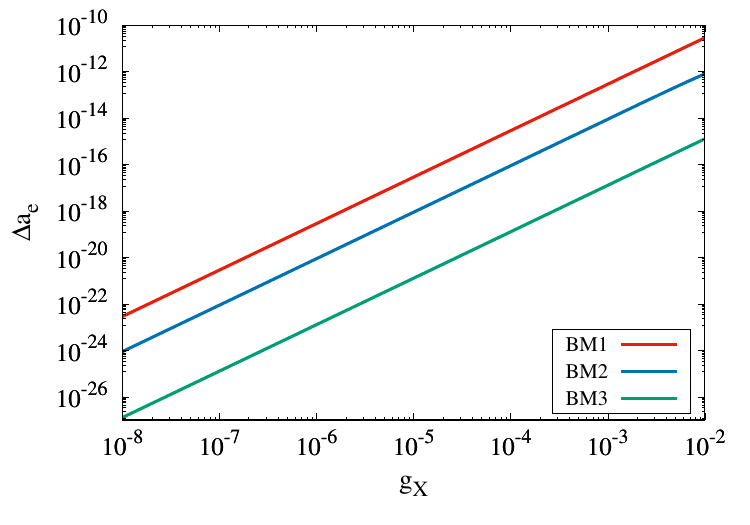}
\caption{Variation of $\Delta\, a_e$ as a function of $g_X$ for BM1~(red), BM2~(blue), and BM3~(green) with the $M_{Z^\prime}$ being fixed at 1 GeV.}
\label{fig:DA_gx}
\end{figure}

Fig.~\ref{fig:DA_gx} shows the variation of $\Delta\,a_e$ with $g_X$ for $M_{Z^\prime}=1$ GeV. The considered BMs, BM1, BM2, and BM3, have been represented in red, blue, and green, respectively. $g_X$ has been varied from $10^{-8}$ to $10^{-2}$ leading to a large variation in $\Delta\,a_e$. Further, one can explicitly understand the leptonic $X$ charge dependence of $\Delta\,a_e$ from Fig.~\ref{fig:DA_gx}. For a given $g_X$, $\Delta\,a_e^{\rm BM1}>\Delta\,a_e^{\rm BM2}>\Delta\,a_e^{\rm BM3}$ as BM1 corresponds to the largest $X_{e_{L,\,R}}$ values among the three BMs. Note that in contrast to the $\rho$ parameter, $\Delta\,a_e$ has no significant dependence on $X_H$. 
\section{Exclusion Limits}
\begin{figure}[!ht]
\centering
\subfloat[\qquad\qquad(a)]{\includegraphics[scale=0.65]{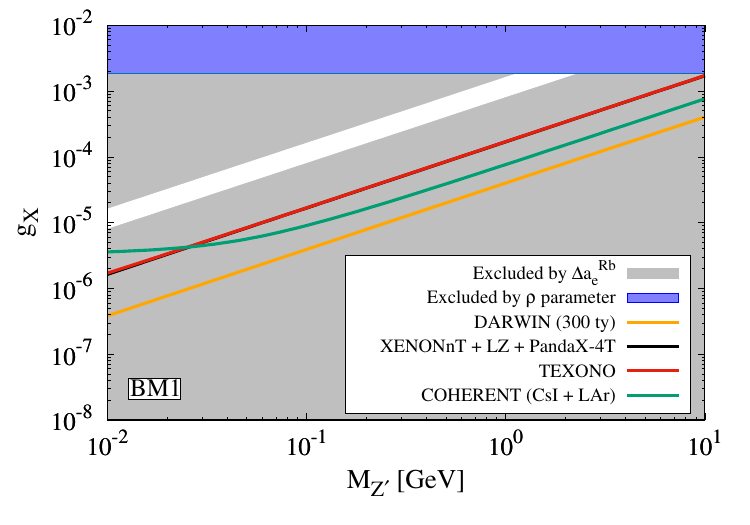}}
\subfloat[\qquad\qquad(b)]{\includegraphics[scale=0.65]{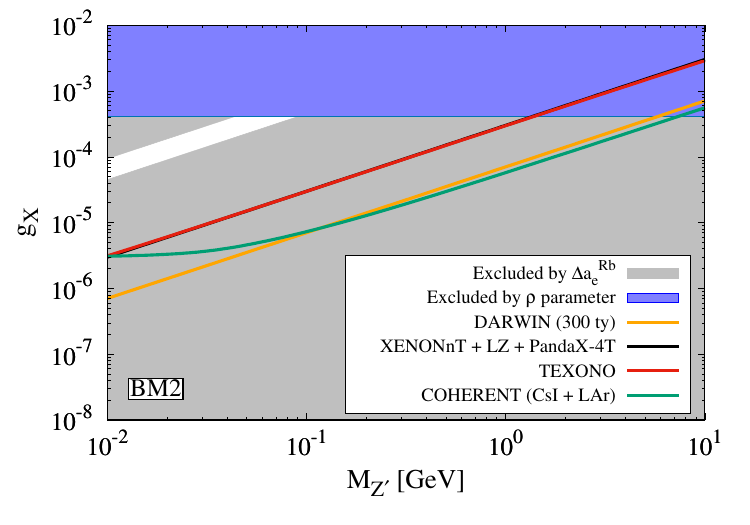}}\\
\subfloat[\qquad\qquad(c)]{\includegraphics[scale=0.65]{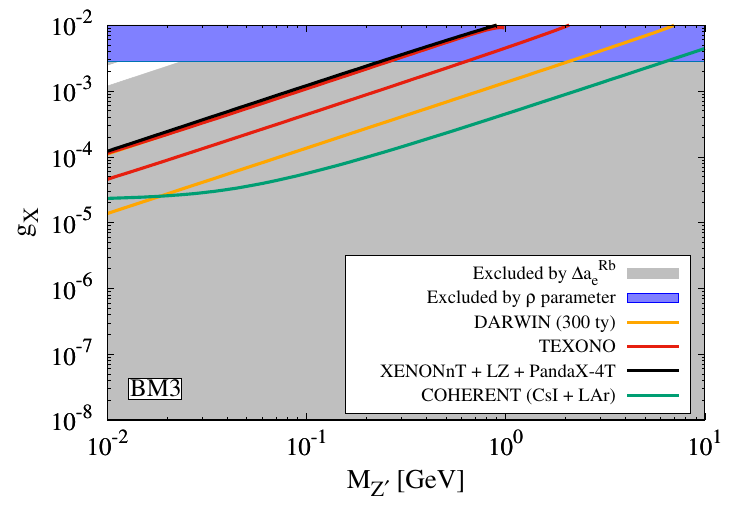}}
\caption{Exclusion limits on the $\left\{M_{Z^\prime},\,g_X\right\}$ parameter space for (a) BM1, (b) BM2, and (c) BM3. The gray shaded region corresponds to $\Delta \, a_e\neq \Delta\,a_e^{\rm Rb}$ while the blue color represents violation of $\rho$ parameter constraint. The colored lines mark the existing experimental bounds on the considered chiral BMs~\cite{Majumdar:2024dms}.}
\label{fig:bounds}
\end{figure}

\noindent
\label{sec:ex}
Fig.~\ref{fig:bounds} displays a compilation of the various experimental bounds on the considered BMs. The blue shade marks the parameter space where $\rho=1.00038 \pm 0.00060$ can't be satisfied for the chiral models. Note that the exclusion limits from the EW global fit are approximately independent of 
$M_{Z^\prime}$ and become more stringent for higher $X_H$ values. To be particular, for BM1, $g_X>1.82\times 10^{-3}$, for BM2, $g_X>4.07\times 10^{-4}$, and for BM3, $g_X>2.86\times 10^{-3}$ are excluded by the EW precision data. The $\rho$ parameter bounds depicted in Fig.~\ref{fig:bounds} are in perfect agreement with Ref.~\cite{Majumdar:2024dms}. The considered flavor-universal chiral models can also generate tree-level corrections to the neutrino-electron and neutrino-nucleus scattering processes. Thus, the experiments searching for elastic neutrino-electron scattering~(E$\nu$ES) and coherent elastic neutrino-nucleus scattering~(CE$\nu$NS) can be pivotal to probe the present BMs. The orange, red, green, and black solid lines represent the experimental constraints corresponding to these neutrino scattering processes from DARWIN~\cite{DARWIN:2020bnc}, TEXONO~\cite{TEXONO:2009knm}, COHERENT~\cite{COHERENT:2021xmm,COHERENT:2020iec}, and the current direct detection experiments~\cite{XENON:2022ltv,LZ:2023poo,PandaX:2024cic}, respectively. The bounds have been borrowed from Ref.~\cite{Majumdar:2024dms}, where the authors have performed a comprehensive detector analysis for the considered BMs. However, note that the detector constraints become weaker for higher $M_{Z^\prime}$ values. For TeV-scale chiral $U(1)_X$ models, the strongest bounds can be found in Ref.~\cite{Prajapati:2024wuu} where the exclusion limits correspond to ATLAS~\cite{ATLAS:2019erb} and CMS~\cite{CMS:2021ctt} searches for $Z^\prime\to$ dilepton resonances and the perturbativity constraint, i.e., $X_{\rm max}g_X\leq \sqrt{4\pi}$ with $X_{\rm max}$ being the largest $U(1)_X$ charge in the theory.

The gray shaded region is the outcome of the present work and highlights the parameter space where $\Delta \,a_e\neq\Delta\,a_e^{\rm Rb}$. It's worth emphasizing that in the chiral $U(1)_X$ extensions of the SM, corrections to the lepton anomalous magnetic moments can't be ignored, and the theory must satisfy the observed experimental data to be a viable one. It has been numerically checked that the allowed parameter spaces of the three BMs are consistent with the latest updates on $(g-2)_\mu$, leading to no additional constraints. However, for the $(g-2)_e$ anomaly, most of the parameter space is futile, leaving only the white strips as $\Delta\,a_e^{\rm Rb}$-allowed regions. With decreasing $U(1)_X$ charges, one needs larger $g_X$ values to satisfy the observed discrepancy, pushing the white strip towards the $\rho$-excluded region. A similar upward shift can also be noticed for the higher $M_{Z^\prime}$ values. However, for all the BMs, the $(g-2)_e$-satisfying parameter spaces have already been ruled out by the existing neutrino scattering experiments. Thus, for the considered chiral models, no parameter space exists where the experimental bounds reported by Ref.~\cite{Majumdar:2024dms} and the $(g-2)_e$ anomaly can be simultaneously satisfied. 
\section{Conclusion}  
\noindent
\label{sec:conc}
In the present paper, the SM gauge group has been extended with an arbitrary $U(1)_X$ symmetry that results in chiral interactions among the SM fermions. Therefore, three SM-singlet right-handed fermions are required to cancel the anomalies and to maintain the renormalizability of the theory. The $X$ charges of all the fields have been fixed through the anomaly cancellation conditions along with the constraint that the SM Yukawa terms must remain invariant under the considered chiral symmetry. Moreover, an SM-singlet complex scalar $\Phi$ has been added to the particle spectrum for breaking the $U(1)_X$ symmetry spontaneously. After SSB, the mass mixing between the EW and $U(1)_X$ gauge bosons results in two massive physical states, $Z$ and $Z^\prime$, inducing a tree-level correction to the $\rho$ parameter. Considering three anomaly-free chiral BMs, the variation of the $\rho$ parameter and the mixing angle $\theta_X$ have been studied as a function of $g_X$, leading to significant exclusion limits on the associated parameter spaces. Note that all the SM fermions being charged under $U(1)_X$, the framework can also induce loop-level corrections to the SM sector through $Z$ and $Z^\prime$. A phenomenologically crucial SM observable is the $(g-2)_e$ --- the discrepancy between its theoretical and experimental values can only be explained in the presence of a BSM contribution. In the considered chiral $U(1)_X$ models, the leading-order corrections to $(g-2)_e$ arise from the $Z$ and $Z^\prime$-induced one-loop diagrams. The effective BSM contribution to $a_e$ being positive over the entire parameter space, $\Delta\,a_e^{\rm Rb}$ has been used to constrain the models. For all the considered BMs, the results show that only a narrow parameter space survives to explain the $(g-2)_e$ anomaly. However, these regions are already excluded by the current detector constraints. Thus, when the $(g-2)_e$ anomaly is considered in complementarity with the E$\nu$ES and CE$\nu$NS experiments, the chiral $U(1)_X$ models can be severely constrained. To be specific, the considered BMs are completely ruled out by the present analysis. However, the bounds on the $\left\{M_{Z^\prime},\,g_X\right\}$ space can be relaxed either with an augmented particle spectrum or by further extending the gauge symmetry. A flavor-specific chiral theory may also improve the scenario. Though the possibilities might be explored in the future, current experiments falsify the considered flavor-blind chiral $U(1)_X$ models as self-sufficient NP theories.
\section*{Acknowledgements}
\noindent
The author gratefully acknowledges R. Srivastava and H. Prajapati for some valuable discussions on the chiral $U(1)_X$ theories.

\bigskip
\small \bibliography{U1_chiral}{}
\bibliographystyle{JHEPCust}    
    
\end{document}